\pdfoutput=1	

\documentclass{elsart}
\usepackage{graphicx}
\usepackage{amssymb}
\usepackage{hyperref}

\begin{document}

\newcommand{\rmi}{\mathrm{i}}
\newcommand{\rmd}{\mathrm{d}}
\newcommand{\bi}[1]{\mathbf{#1}}
\newcommand{\ul}[1]{\underline{#1}}
\newcommand{\p}{^\prime}

\begin{frontmatter}

\title{Geometric limits to geometric optical imaging with infinite, planar, non-absorbing sheets}

\author{Johannes Courtial}

\address{Department of Physics \& Astronomy, Faculty of Physical Sciences, University of Glasgow, Glasgow G12~8QQ, UK}

\ead{j.courtial@physics.gla.ac.uk}

\begin{abstract}
Wave optics can limit the ways in which optical components can change light-ray fields.
Optical components called METATOYs trade in the continuity of the phase fronts and the precision to which they change light-ray fields in return for additional possibilities when changing light-ray fields.
Now only geometry limits the possible mappings between the positions of an object and its geometric image.
Here I study such limitations for the case of an infinite, planar, non-absorbing sheet that images the entire three-dimensional space.
The most general case of such a sheet is equivalent to a thin lens with different object- and image-sided focal lengths.
Special cases include ordinary thin lenses, confocal lenslet arrays, and negative refraction with $n_2 = -n_1.$
\end{abstract}


\end{frontmatter}

\section{Introduction}
Motivated by a number of similarities with metamaterials \cite{Smith-et-al-2004}, we recently started to investigate 2-dimensional arrays of miniaturized optical components.
If the components are telescopic (in the sense that they transform bundles of parallel light rays into other bundles of light rays that can have different directions but are again parallel), then the light-ray-direction change is independent of the precise position where a light ray hits an optical component.
Examples of telescopic components we have examined in arrays include Dove prisms \cite{Hamilton-Courtial-2008a}, pairs of Dove prisms that are rotated with respect to each other \cite{Courtial-Nelson-2008,Hamilton-et-al-2009}, simple telescopes in the form of confocal pairs of spherical lenses (lenses that share a common focal plane) \cite{Courtial-2008a} or generalized telescopes made from confocal pairs of inclined elliptical lenses \cite{Hamilton-Courtial-2009b}.
In addition to changing the direction of transmitted light rays, the telescopic components offset transmitted light rays, but in such a way that miniaturizing the components also miniaturizes the offset.
For many visual applications, the offset can be made so small that it can be neglected.
The result is a thin transparent sheet with a homogeneous appearance that changes the direction of transmitted light rays in unusual ways.
We call such sheets \ul{meta}ma\ul{t}erials f\ul{o}r ra\ul{y}s (METATOYs) \cite{Hamilton-Courtial-2009}.

Optical instruments are traditionally designed to work in air or at least with an isotropic medium at either end (the oil immersion microscope is an example of the latter, but not the former), and such that a non-pathological (specifically continuous and differentiable ``almost everywhere'') incident phase front is still non-pathological as it exits.
In the ray-optics limit of wave optics, this limits the possible configurations of light-ray fields:
as the light-ray direction is given by the normal to the phase front (that is, the gradient of the phase), and as the curl of any gradient field has to vanish, the curl of the light-ray-direction field also has to vanish to ensure that the corresponding wave field is non-pathological \cite{Hamilton-Courtial-2009}.\footnote{This argument glosses over a subtle point, namely how the length of the phase-gradient vector relates to the ``length'' of the corresponding light-ray-direction vector, which is discussed in Ref.\ \cite{Hamilton-Courtial-2009}.}
For our purposes, this means that a light-ray direction is not independent of neighbouring light-ray directions.

Each miniaturized optical component in a METATOY performs a piecewise transformation of the direction of the phase front, and therefore of the local light-ray direction; between components, the phase is allowed to be discontinuous.
METATOYs thereby compromise:  they sacrifice global continuity of the phase fronts but gain complete independence of the local light-ray direction.
Detailed computer simulations have demonstrated that METATOY refraction can be made to work well enough to show the expected visual effects \cite{Hamilton-Courtial-2008a,Courtial-Nelson-2008,Hamilton-et-al-2009,Courtial-2008a,Hamilton-Courtial-2009b,Hamilton-Courtial-2008c}, and very recently results from these simulations have started to be confirmed experimentally \cite{Blair-et-al-2009}.

Perhaps our favourite example of a METATOY is a sheet formed from the aforementioned pairs of rotated Dove prisms, which can rotate the local light-ray direction around the sheet normal \cite{Hamilton-et-al-2009}.
It was this local light-ray direction that we used to demonstrate that METATOYs create light-ray fields that are not limited by the wave-optical requirement to have vanishing curl \cite{Hamilton-Courtial-2009}.
That local light-ray rotation violates wave-optical principles is further demonstrated by forced application of the wave-optically motivated Fermat's principle \cite{Born-Wolf-1980-Fermat} to local light-ray rotation, resulting a formal description in terms of a complex refractive-index ratio \cite{Sundar-et-al-2009}.

What Fermat's principle is for refraction, the principle of equal optical path \cite{Born-Wolf-1980-principle-of-equal-optical-path} is for imaging.
This principle is the basis for a number of theorems about imaging \cite{Born-Wolf-1980-perfect-imaging}, but as this principle is again wave-optically motivated and does not apply to METATOY refraction, imaging with METATOYs is not subject to those theorems and corresponding restrictions.
This means that METATOY imaging might well offer new possibilities.

Here I consider an idealized METATOY in the form of an infinite, planar, non-absorbing sheet that images all of three-dimensional space.
Point light sources at three-dimensional (object) positions $P_i$ are imaged into corresponding image positions $P_i\p$.
I show that a combination of an idealized lens and confocal lenslet arrays \cite{Courtial-2008a} performs the most general mapping between the object and image positions.

The argument makes the following assumptions:
\begin{enumerate}
\item The optical system is an infinitely thin, infinite, planar, and non-absorbing sheet that changes only the direction of transmitted light rays, and so does not offset them.
\item If the sheet geometrically images a point object at position $P_1$, all light rays that originate at $P_1$ and pass through the sheet intersect in the image position, $P_1\p$.
\footnote{Both object and image can be real or virtual.  In that case, it is not the actual light rays, but their straight-line continuations, that intersect at the object or image position.}
\item
On either side of the sheet, all light rays travel in straight lines, as is the case in homogeneous media.
\end{enumerate}

The limits derived here are all based on the following, very simple, observation.
Two light rays that enter the sheet along the same trajectory will leave it along the same trajectory.
Specifically, two light rays that originate at different point objects $P_1$ and $P_2$ and that enter the sheet along the straight line through both point objects will, after passage through the sheet, follow a common straight-line trajectory.
The corresponding images, $P_1\p$ and $P_2\p$, provided they exist, must therefore lie on the straight line defined by this trajectory.

\section{\label{geometry-section}Derivation of the imaging equations}
Let us start with two object-image pairs, $(P_1, P_1\p)$ and $(P_2, P_2\p)$.
Just like the straight-line light path through the objects and that through the images in any other pair of object-image pairs, the line $P_1 P_2$, by which we mean the straight line through $P_1$ and $P_2$ or its continuation, and the line $P_1\p P_2\p$ have to intersect in a point in the sheet plane.
We call this point $I$.

\begin{figure}
\begin{center}
\includegraphics{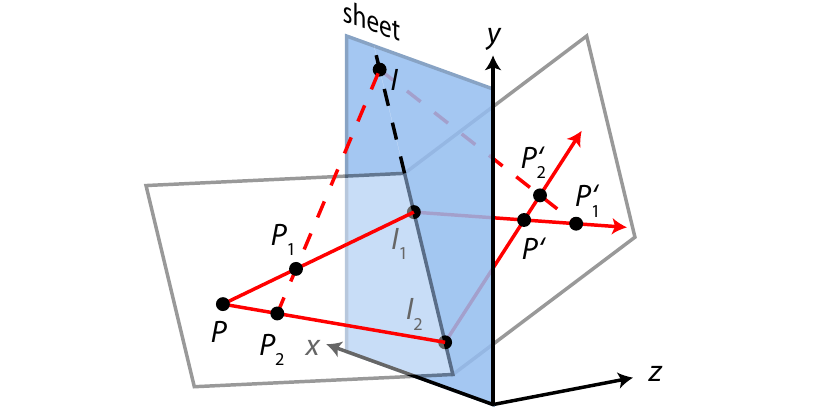}
\end{center}
\caption{\label{construct-PPrime-figure}(Color online)
Construction of the image position $P\p$ corresponding to the object position $P$, given two pairs of object and image positions, $(P_1,P_1\p)$ and $(P_2,P_2\p)$.
$P\p$ lies at the intersection of two light rays (solid arrows), both originating at $P$:  one that passes through $P_1$ and is then refracted by the sheet, and another that passes through $P_2$ before being refracted.
The pairs of object and image positions $(P_1,P_1\p)$ and $(P_2,P_2\p)$ themselves have to lie on the path of a single light ray (dashed line).}
\end{figure}

To ensure that the lines $P_1 P_2$ and $P_1\p P_2\p$ intersect in a point, we define the points $P_2$ and $P_2\p$ in terms of the intersection point, $I$, and respectively $P_1$ and $P_1\p$:
\begin{eqnarray}
P_2 &=& I + a (P_1 - I), \\
P_2\p &=& I + b (P_1\p - I).
\end{eqnarray}
We describe the points $P_1$, $P_1\p$, and $I$ in terms of their coordinates in a Cartesian coordinate system whose $(x,y)$ plane is in the sheet plane:
\begin{equation}
P_1 = \left( \begin{array}{c} x_1 \\ y_1 \\ z_1 \end{array} \right), \quad
P_1\p = \left( \begin{array}{c} x_1\p \\ y_1\p \\ z_1\p \end{array} \right), \quad
I = \left( \begin{array}{c} x_{I} \\ y_{I} \\ 0 \end{array} \right).
\end{equation}

We want to understand the implications for the position of a third object-image pair, $(P,P\p)$.
First we check whether it is possible that every point $P$ is actually imaged.
The light ray that passes through $P$ and $P_1$, after refraction at point $I_1$ on the sheet, has to pass through both $P\p$ and $P_1\p$ (not necessarily in that order).
Similarly, the light ray passing through $P$ and $P_2$ has to pass, after refraction at point $I_2$, through both $P\p$ and $P_2\p$.
In order for this to be possible, the lines $I_1 P_1\p$ and $I_2 P_2\p$ must actually intersect.
For $P$ to be imaged by the sheet, consistent with the object-image pairs $(P_1,P_1\p)$ and $(P_2,P_2\p)$, the condition that the points $I_1$, $I_2$, $P_1\p$ and $P_2\p$ lie in the same plane therefore has to be satisfied.
This is the case, as the lines $I_1 I_2$ and $P_1\p P_2\p$ intersect (in the point where the lines through $P_1$ and $P_2$ also intersects, namly $I$, Fig.\ \ref{construct-PPrime-figure}).
$P\p$ will therefore exist, and lie in the plane defined by the line through the $I$s and that through the $P\p$s.

We calculate the position of $P\p$ by calculating $I_1$ (as the intersection of the line $P_1 P$ and the sheet), $I_2$ (the intersection of the line $P_2 P$ and the sheet), and then intersecting the lines $I_1 P_1\p$ and $I_2 P_2\p$.
With
\begin{equation}
P = \left( \begin{array}{c} x_{P} \\ y_{P} \\ -o \end{array} \right)
\end{equation}
($o$ is the object distance, that is, distance in front of the sheet),
the result is
\begin{eqnarray}
P\p = \frac{1}{(a-b) o + (a - a b) z_1}
\left( \begin{array}{c}
(a - a b) (x_P z_1 - x_1 o) + (b - a b) x_1\p o \\ 
(a - a b) (y_P z_1 - y_1 o) + (b - a b) y_1\p o \\ 
(b - a b) z_1\p o
\end{array} \right).
\label{Pp-equation}
\end{eqnarray}

Before developing a more compact form of the mapping described by Eqn (\ref{Pp-equation}), we briefly discuss a few details of this mapping.
\begin{itemize}
\item Eqn (\ref{Pp-equation}) describes a mapping, defined by two object-image pairs, $(P_1,P_1\p)$ and $(P_2,P_2\p)$, from an object position to a corresponding image position.
This can be used to define a third object-image pair, which we call $(P_3,P_3\p)$.
It is relatively straightforward to show that the mapping defined by one of the original object-image pairs, say $(P_1,P_1\p)$, and the new object-image pair $(P_3,P_3\p)$, is the same as that defined by the original two object-image pairs.
In this sense, our approach is self-consistent.
\item So far we have \emph{assumed} that $P$ is imaged.
What we have shown so far is that the two light rays that respectively pass through $P$ and $P_1$ and through $P$ and $P_2$, and that are then transmitted through the sheet, have to intersect at the position $P\p$.
What we have not shown so far is that \emph{all} light rays that pass through $P$, after transmission through the sheet, intersect again at $P\p$, so we have not shown that the sheet geometrically images $P$ to $P\p$.
Now we are in a position to do so.
If we want to show that light rays that leave $P$ in a particular direction and pass through the sheet intersect at $P\p$, we simply pick an object position $P_3 \neq P$ on the light ray leaving $P$ in the direction of interest and construct the position of its image, $P_3\p$.
Because the mapping defined by any one of the original object-image pairs and by $(P_3,P_3\p)$ is the same as that defined by the original two object-image pairs, the light ray through $P$ and $P_3$ has to intersect $P\p$.
As this is true for any object position $P_3$, \emph{all} light rays passing through $P$ and the sheet intersect at $P\p$, so $P\p$ is the image of $P$.
\item The two object-image pairs $(P_1,P_1\p)$ and $(P_2,P_2\p)$ define a one-to-one mapping for all object positions other than those on the line $P_1 P_2$; for object positions on the line $P_1 P_2$, our construction merely restricts the image position to the line $P_1\p P_2\p$.
By constructing a third object-image pair $(P_3,P_3\p)$ (whereby $P_3$ must not lie on the line $P_1 P_2$), we can then construct a unique image position of any point on the line $P_1 P_2$ as that constructed from the two object-image pairs $(P_1,P_1\p)$ and $(P_3,P_3\p)$ (or, equivalently, $(P_2,P_2\p)$ and $(P_3,P_3\p)$).
\end{itemize}
In the following we express the mapping from $P$ to $P\p$ in a more compact form.

First we calculate, in analogy to calculating the image-sided focal point, the image position $P\p$ in the limit of infinite object distance, $o$, and finite transverse object coordinates, $x_P$ and $y_P$.
The result,
\begin{equation}
G = \lim_{o \rightarrow \infty} P\p = 
\left( \begin{array}{c}
  \left[ (a - a b) x_1 - (b - a b) x_1\p \right] / (a-b) \\
  \left[ (a - a b) y_1 - (b - a b) y_1\p \right] / (a-b) \\
  g
\end{array} \right),
\label{G-equation}
\end{equation}
where
\begin{equation}
g = - \frac{b - a b}{a-b} z_1\p,
\label{g-equation}
\end{equation}
is independent of $x_{P}$ and $y_{P}$.
It therefore has all the properties of an image-sided focal point.
We call its distance behind the sheet, $g$, the image-sided focal length.
Similarly, we can define and calculate an object-sided focal point, $F$.
It turns out that the $x$ and $y$ coordinates of $F$ are the same as those of $G$, and that the object-sided focal length, $f$ (that is, the negative $z$ component of $F$), is
\begin{equation}
f = \frac{a-a b}{a-b} z_1.
\label{f-equation}
\end{equation}

Secondly, we use more suitable coordinates. 
The fact that the two focal points have the same $x$ and $y$ components suggests that the sheet has an optic axis that passes through the two focal points.
This in turn suggests making the optic axis the $z$ axis.
In these new coordinates, $P$ and $P\p$ are
\begin{equation}
P = \left( \begin{array}{c} x \\ y \\ -o \end{array} \right), \quad
P\p = \left( \begin{array}{c} x\p \\ y\p \\ i \end{array} \right).
\end{equation}
When formulated in these new coordinates and in terms of the focal lengths $f$ and $g$, the equations for $P\p$, equations (\ref{Pp-equation}), become
\begin{equation}
\frac{x\p}{x} = \frac{f}{f - o}, \quad
\frac{y\p}{y} = \frac{f}{f - o}, \quad
\frac{f}{o} + \frac{g}{i} = 1.
\label{xyz-equation}
\end{equation}

These equations are the main result of this paper.
They describe the most general imaging performed by a thin, planar, transparent sheet that is embedded in a homogeneous medium and that images all of three-dimensional space.

\section{Lenses, confocal lenslet arrays, and negative refraction}
The first two equations in (\ref{xyz-equation}) describe the relationship between the $x$ and $y$ components of the object and image positions.
They describe the ratio of the transverse image and object coordinates, which is the transverse magnification.
This transverse magnification is the same in the $x$ and $y$ directions, and it is dependent only on $f$ and the object distance; specifically, it is independent of $g$.
It is also the same as in the case of a thin lens of focal length~$f$.

The third equation of (\ref{xyz-equation}) describes the relationship between object and image distance.
It is almost, but not quite, the same as that for a thin lens of (object- and image-sided) focal length $f$, which can be written as
\begin{equation}
\frac{f}{o} + \frac{f}{i_\mathrm{lens}} = 1.
\end{equation}
The difference is simple scaling of image space in the axial direction:
for the same object distance and (object-sided) focal length, the image distances in the general case and in the lens case scale with their image-sided focal lengths:
\begin{equation}
\frac{i}{i_\mathrm{lens}} = \frac{g}{f}.
\end{equation}

The stretching of image space in the $z$ direction can be seen particularly clearly in the case of infinite focal lengths $f$ and $g$.
At first, this sounds like the case of a thin lens with zero focussing power, and indeed no focussing takes place.
However, this does not take into account the fact that $f$ and $g$ can tend to infinity in specific ways:
if the ratio of $f$ and $g$ is fixed before $f$ -- and with it $g$ -- is sent to infinity, so
\begin{equation}
\frac{g}{f} = \eta, \quad f \rightarrow \pm \infty,
\label{eta-equation}
\end{equation}
then the third equation of (\ref{xyz-equation}) becomes
\begin{equation}
i = -\eta o.
\end{equation}
This is pure scaling of the axial direction of image space relative to that of object space, relative to the sheet plane, by a factor $\eta$. 
This is precisely the geometric imaging performed by two confocal lenslet arrays, one with focal length $f_1$, the other with focal length $f_2$, with a focal-length ratio~\cite{Courtial-2008a}
\begin{equation}
\frac{f_2}{f_1} = -\eta.
\end{equation}
A thin lens of focal length $f$, immediately followed by two confocal lenslet arrays with a focal-length ratio
\begin{equation}
\eta = \frac{g}{f},
\end{equation}
therefore images according to equations (\ref{xyz-equation}).

A particular example of the scaling of the axial direction is the case
\begin{equation}
i = o,
\end{equation}
which corresponds to a $z$ scaling factor $\eta = -1$.
According to equation (\ref{eta-equation}), this is the case in the limit of $f$ and $g$ respectively being positive and negative infinity (or vice versa), so
\begin{equation}
g = -f \rightarrow \pm \infty.
\end{equation}
As the object- and image-sided focal lengths are measured in opposite directions from the sheet, this corresponds to both focal points being at an infinite distance on the same side of the sheet.
In the case of such a sheet, the image distance equals the object distance (for any object distance), but as the object and image distances are measured in opposite directions from the sheet, the position of the image is that of the object, mirrored with respect to the sheet plane.
This is precisely the geometric imaging due to negative refraction at a planar interface between media with refractive indices $n$ and $-n$, respectively~\cite{Pendry-2000}.

\section{Conclusions}
In the end, imaging with infinite, planar, non-absorbing sheets turns out to be only a slight generalization of imaging with a perfect thin lens:  the generalization is that the object- and image-sided focal lengths can be different.
Nevertheless, this opens up new possibilities, such as a description of negative refraction and scaling of the axial direction of image space relative to that of object space.
These possibilities can be realized with combinations of lenses and confocal lenslet arrays~\cite{Courtial-2008a}.

\section*{Acknowledgments}

Thanks to Bhuvanesh Sundar and Alasdair C.\ Hamilton for helpful comments.
J.\ Courtial is a Royal Society University Research Fellow.


\begin{thebibliography}{10}
\newcommand{\enquote}[1]{``#1''}

\bibitem{Smith-et-al-2004}
D.~R. Smith, J.~B. Pendry, and M.~C.~K. Wiltshire, \enquote{{Metamaterials and
  Negative Refractive Index},} Science \textbf{305}, 788--792 (2004).

\bibitem{Hamilton-Courtial-2008a}
A.~C. Hamilton and J.~Courtial, \enquote{Optical properties of a {D}ove-prism
  sheet,} J. Opt. A: Pure Appl. Opt. \textbf{10}, 125302 (2008).

\bibitem{Courtial-Nelson-2008}
J.~Courtial and J.~Nelson, \enquote{Ray-optical negative refraction and
  pseudoscopic imaging with {D}ove-prism arrays,} New J. Phys. \textbf{10},
  {023028} (2008).

\bibitem{Hamilton-et-al-2009}
A.~C. Hamilton, B.~Sundar, J.~Nelson, and J.~Courtial, \enquote{Local light-ray
  rotation,} J. Opt. A: Pure Appl. Opt. (in press) (2009).

\bibitem{Courtial-2008a}
J.~Courtial, \enquote{Ray-optical refraction with confocal lenslet arrays,} New
  J. Phys. \textbf{10}, 083033 (2008).

\bibitem{Hamilton-Courtial-2009b}
A.~C. Hamilton and J.~Courtial, \enquote{Generalized refraction using lenslet
  arrays,} J. Opt. A: Pure Appl. Opt. \textbf{11}, 065502 (2009).

\bibitem{Hamilton-Courtial-2009}
A.~C. Hamilton and J.~Courtial, \enquote{Metamaterials for light rays: ray
  optics without wave-optical analog in the ray-optics limit,} New J. Phys.
  \textbf{11}, 013042 (2009).

\bibitem{Hamilton-Courtial-2008c}
A.~C. Hamilton and J.~Courtial, \enquote{Imaging with parallel ray-rotation
  sheets,} Opt. Express \textbf{16}, 20826--20833 (2008).

\bibitem{Blair-et-al-2009}
M.~Blair, L.~Clark, E.~A. Houston, G.~Smith, J.~Leach, A.~C. Hamilton, and
  J.~Courtial, \enquote{Experimental demonstration of a
  light-ray-direction-flipping {METATOY} based on confocal lenticular arrays,}
  arXiv:0902.3192 [physics.optics] (2009).

\bibitem{Born-Wolf-1980-Fermat}
M.~Born and E.~Wolf, \emph{Principles of Optics} (Pergamon Press, Oxford,
  1980), chap. 3.3.2.

\bibitem{Sundar-et-al-2009}
B.~Sundar, A.~Hamilton, and J.~Courtial, \enquote{Fermat's principle with
  complex refractive indices and local light-ray rotation,} Opt. Lett.
  \textbf{34}, 374--376 (2009).

\bibitem{Born-Wolf-1980-principle-of-equal-optical-path}
M.~Born and E.~Wolf, \emph{Principles of Optics} (Pergamon Press, Oxford,
  1980), chap. 3.3.3.

\bibitem{Born-Wolf-1980-perfect-imaging}
M.~Born and E.~Wolf, \emph{Principles of Optics} (Pergamon Press, Oxford,
  1980), chap. 4.2.

\bibitem{Pendry-2000}
J.~B. Pendry, \enquote{Negative refraction makes a perfect lens,} Phys. Rev.
  Lett. \textbf{85}, 3966--3969 (2000).

\end{thebibliography}

\end{document}